# A Sequential Density-Based Empirical Likelihood Ratio Test for Treatment Effects

Li Zou[1], Albert Vexler[2], Jihnhee Yu[2], and Hongzhi Wan[2]

In health-related experiments, treatment effects can be identified using paired data that consist of pre- and post-treatment measurements. In this framework, sequential testing strategies are widely accepted statistical tools in practice. Since performances of parametric sequential testing procedures vitally depend on the validity of the parametric assumptions regarding underlying data distributions, we focus on distribution-free mechanisms for sequentially evaluating treatment effects. In fixed sample size designs, the density-based empirical likelihood (DBEL) methods provide powerful nonparametric approximations to optimal Neyman-Pearson type statistics. In this article, we extend the DBEL methodology to develop a novel sequential DBEL testing procedure for detecting treatment effects based on paired data. The asymptotic consistency of the proposed test is shown. An extensive Monte Carlo study confirms that the proposed test outperforms the conventional sequential Wilcoxon signed-rank test across a variety of alternatives. The excellent applicability of the proposed method is exemplified using the ventilator-associated pneumonia study that evaluates the effect of Chlorhexidine Gluconate treatment in reducing oral colonization by pathogens in ventilated patients.

## 1. INTRODUCTION

In health-related studies, in order to test for treatment effects oftentimes investigators sequentially collect paired data that consist of pre- and post-treatment measurements.[1] The proposed method in this article is motivated by the following example. The oral cavity, especially dental plaque biofilms, may be colonized by potential respiratory pathogens (PRPs) in mechanically-ventilated (MV), intensive care unit (ICU) patients. Thus, by improving oral hygiene for the MV-ICU patients, we may prevent ventilator-associated pneumonia (VAP). One of the primary goals of the VAP study [2] was to evaluate the effect of Chlorhexidine Gluconate (CHX) treatment, a cationic chlorophenyl bis-biguanide antiseptic, in reducing oral colonization by pathogens in MV-ICU patients. The trial sequentially enrolled ventilated patients who were admitted to a trauma ICU of the Erie County Medical Center (ECMC). During this study, pre- and post-CHX treatment measurements of amount of aggregated bacteria (S. aureus, P. aeruginosa, Acinetobacter sps.) and enteric organisms (Klebsiella pneumoniae, Serratia marcescens, Enterobacter sps., Proteus mirabilis, E. coli) were recorded from each patient. We aim to develop a novel sequential methodology that requires substantially fewer numbers of patients to make a conclusion regarding CHX treatment effect based on the data from the VAP study.

[1]*Department of Statistics and Biostatistics, California State University, East Bay, Hayward, CA 94542*
[2]*Department of Biostatistics, The State University of New York at Buffalo, Buffalo, NY 14214*
*Correspondence to*: *Albert Vexler, Department of Biostatistics, The State University of New York at Buffalo*
*E-mail: avexler@buffalo.edu*

It is desirable for investigators to detect treatment effects as early as possible, since the following two reasons are in effect: 1) The ethical reason: the ongoing clinical trials can have early termination when the significant superiority of the new therapy is statistically proven. In this case, stopping early enables other patients to begin receiving the superior treatment sooner. [3] 2) The efficiency reason: early termination of a trial yields savings in sample size. Thus, the sequential testing strategies can save resources and time. In these contexts, sequential statistical methods are important and frequently employed tools in practice. The statistical literature has dealt extensively with both the theoretical and applied aspects of various sequential statistical designs. [4-8]

Commonly, to implement sequential statistical procedures, parametric assumptions regarding the underlying data distribution are stated. Performances of parametric sequential testing procedures strongly depend on the assessment of the parametric assumptions of data distributions. Retrospective studies are generally based on already collected datasets or combining existing pieces of data. [9] In contrast with the analysis of data obtained retrospectively, we have the following problems related to sequential analysis. First, it is difficult to specify the parametric distribution form of the underlying data before data points are observed. Second, even if we have strong reasons to assume the parametric form of the data distribution, it will be extremely difficult, for example, to test the corresponding parametric assumptions after the execution of sequential procedures. Since sequential tests are assumed to be based on random number of observations, data obtained after sequential analyses cannot be evaluated for goodness-of-fit using the conventional retrospective tests. [10] Toward this end, in this article we focus on an efficient nonparametric sequential approach.

The modern theory of sequential analysis originates from the researches of Barnard [11] and Wald [12]. In particular, the method of the sequential probability ratio test (SPRT) has been the predominant influence of the subsequent developments in the area. Note that since analytical forms of underlying data distributions are not completely specified, one cannot employ the most powerful SPRT testing strategy via the Neyman-Pearson concept. [13] As an alternative, Miller [14] proposed a nonparametric sequential signed-rank test (SSRT), employing a Wilcoxon type test statistic. Wilcoxon type tests are commonly used to detect treatment effects based on paired data in fixed sample size designs. [15-17] In accordance with the repeated significance test, [18] the Miller's approach fixes the total number of data points to be $N$ and consists of a series of independent observations. The SSRT is performed after each data point, and stops either when the maximum sample size is $N$ or the test statistic rejects the null hypothesis. In the settings of the SSRT, we introduce a simple sequential test with high and stable power for detecting treatment effects over a broad spectrum of alternatives based on paired data. Oppegaard et al. [1] recently applied a two-sample sequential Wilcoxon test in order to evaluate the difference in preoperative cervical dilation before hysteroscopy between postmenopausal women who receive vaginal misoprostol and postmenopausal women who receive vaginal placebo.

In this article, we present a novel nonparametric sequential testing procedure based on paired data, employing the data-driven likelihood ratio principle. The proposed method uses a nonparametric testing strategy that approximates corresponding optimal parametric Neyman-Pearson type statistics. The density-based empirical likelihood (EL) approach [19,20] is modified and extended to be applied in the sequential setting. The proposed technique can be employed in the context of a single-arm trial, when a sample of individuals with the specified medical condition is given the study therapy and then followed over time to measure their outcomes.

The statistical literature has shown that the EL methodology is a very powerful inference tool in various nonparametric settings. [21-24] The conventional EL concept is outlined in Section 2.3. To the best of our knowledge, the conventional EL concept and the density-based empirical likelihood (DBEL) methodology have not been extensively studied in the statistical sequential literature. The proposed method is distribution-free, robust to underlying model settings and highly efficient. We evaluate the performance of the new sequential DBEL method and the SSRT tests in terms of the statistical power and the average sample number (ASN). The ASN, by definition, means the expected value of the sample size for making decisions. The proposed test demonstrates significantly higher power and smaller ASN than those of the SSRT test across a variety of alternatives treated in an extensive MC study. In this article, we establish the asymptotic consistency of the proposed test. It is clear that in general nonparametric settings there are no most powerful statistical mechanisms. Thus, it is very important to consider various reasonable and efficient distribution-free schemes in the framework of sequential analysis.

This article is organized as follows. Section 2 outlines the classical SSRT test and introduces the proposed method. The development of the new test statistic is presented and its asymptotic properties are derived. In Section 3, an extensive MC study is conducted to evaluate the proposed method. In Section 4, the applicability of the proposed method is illustrated via a clinical trial study related to VAP. In this study, we detect treatment effect of CHX in ventilated patients by using the proposed method, while the SSRT test fails to show the corresponding statistical significance. In Section 5, we provide concluding remarks. The online supplementary material consists of the proof of the theoretical results presented in this article, the R-codes to implement the proposed method and the scatterplot that depicts the data considered in Section 4.

## 2. SEQUENTIAL TESTING METHODS

### 2.1 Hypothesis Setting

In this section, we formalize the statement of the problem, describe the conventional methodology and introduce the novel approach. Let $(X_i, Y_i)$, $i=1,2,\ldots$, denote sequentially surveyed independent and identically distributed (*i.i.d.*) pairs of observations within a subject $i$, where $X_i$ represents the pre-treatment measurement and $Y_i$ represents the post-treatment measurement. Assume the maximum number of subjects allowed in the experiment is $N$. Define $Z_i = X_i - Y_i$, $1 \leq i \leq N$. The nonparametric statistical literature [17] tends to associate the problem of detecting treatment effects with the problem of testing the following null hypothesis:

$$H_0 : F = F_{H_0}, F_{H_0}(u) = 1 - F_{H_0}(-u), \text{ for all } -\infty < u < \infty \text{ versus}$$

$$H_1 : F = F_{H_1}, F_{H_1}(u) \neq 1 - F_{H_1}(-u), \text{ for some } -\infty < u < \infty, \quad (2.1)$$

where $F(u) = \Pr(Z \leq u)$ is an unknown distribution function of i.i.d. $Z_i, i \geq 1$.

## 2.2 Sequential Signed-rank Test for Treatment Effect based on Paired Data

In this section, we outline the commonly used SSRT test for hypothesis (2.1) in practice.[14] Let $R_{ni}$, $i=1,\ldots,n$, be the rank of $|Z_i|$ in $|Z_1|,\ldots,|Z_n|$. The SSRT test statistic is $SR_n = \sum_{i=1}^{n} I(Z_i \geq 0) R_{ni}$, where $I(\cdot)$ is the indicator function. In order to sequentially test for the null hypothesis (2.1), Miller[14] applied the following stopping rule

$$\tau = \min\{n : n \geq 1, TS_n \geq z_{\alpha,N}\}, \quad (2.2)$$

where $TS_n = |SR_n - n(n+1)/4|(n(n+1)(2n+1)/24)^{-1/2}$, $\alpha$ is the pre-specified significance level and $z_{\alpha,N}$ is the critical value associated with $N$ and $\alpha$. The decision making policy consists of the following algorithm: 1) if, for the first time, for some $n (\leq N)$, $TS_n$ exceeds $z_{\alpha,N}$, we have $\tau = n$ and the experiment is terminated at that stage along with the decision to reject $H_0$; 2) if, no such $n$ exists, the null hypothesis is not rejected along with the termination of the experiment at the target maximum sample size $N$.

The critical value $z_{\alpha,N}$ can be determined using the following scheme. Define $W_N = \max_{1 \leq n \leq N}\{TS_n\}$. Let $\Pr_{H_k}$ be the probability measure corresponding to the hypothesis $H_k$ ($k=0$, 1). Thus, by definition (2.2), $z_{\alpha,N}$ is the upper $\alpha$-percentile of the distribution of $W_N$ satisfying $\Pr_{H_0}\{W_N \geq z_{\alpha,N}\} = \alpha$. Note that, under $H_0$, the distribution function of $W_N$ is data-distribution-free, since $\Pr_{H_0}\{W_N \leq z_{\alpha,N}\} = \Pr_{H_0}\{TS_1 \leq z_{\alpha,N},\ldots,TS_N \leq z_{\alpha,N}\}$ and the joint distribution function of $\{TS_n : 1 \leq n \leq N\}$ does not depend on the underlying data distribution function of $Z_1,\ldots,Z_N$. The structure of the statistic $W_N$ is complicated. In practice, one can use the MC methodology in order to evaluate $\Pr_{H_0}\{W_N \geq z_{\alpha,N}\}$ and compute critical values of $z_{\alpha,N}$ for various choices of $N$ and $\alpha$.[14]

## 2.3 Sequential Density-based Empirical Likelihood Ratio Test based on Paired Data

In this section, we develop the DBEL based method for sequentially detecting treatment effects. We begin with considerations related to a retrospective statement of the testing problem, i.e. the sample size is fixed to be $n$.

***The EL approach:*** In order to outline the conventional EL concept, we assume that $U_1,...,U_n$ are *i.i.d.* data points with mean $E(U)$. Consider the commonly used EL ratio test for the null hypothesis $H_0: E(U) = \theta_0$ vs. $H_1: E(U) \neq \theta_0$, where $\theta_0$ is known. In this case, the EL function has the form $EL = \prod_{i=1}^{n} p_i$, where $p_i$'s are the probability weights. Under $H_0$, values of $p_i$'s can be derived by maximizing the *EL* function under the empirical constraints $\sum_{i=1}^{n} p_i = 1$ and $\sum_{i=1}^{n} U_i p_i = \theta_0$. Under the alternative hypothesis $H_1$, the *EL* function is given by $EL = n^{-n}$, since $\prod_{i=1}^{n} p_i$ is maximized by $p_i = n^{-1}$ (*i*=1,...,*n*), when the only constraint $\sum_{i=1}^{n} p_i = 1$ is in effect. In this framework, we reject $H_0$ for large values of $-2\sum_{i=1}^{n} \log(np_i)$ that presents $-2\log\{(EL \text{ under } H_0)/(EL \text{ under } H_1)\}$. This methodology is well developed when data is collected retrospectively, i.e. $n$ is fixed.

***The DBEL approach:*** Motivated by the well-known Neyman-Pearson lemma, Vexler and Gurevich [25] used the EL concept to develop the distribution-free density-based EL (DBEL) methodology for approximating parametric likelihood ratio type statistics. The DBEL method proposes to consider the likelihood in the form

$$DBEL_f = \prod_{i=1}^{n} f(U_i) = \prod_{i=1}^{n} f_i, \quad f_i = f(U_{(i)}),$$

where $f(\cdot)$ is a density function of $U_1,...,U_n$, and $U_{(1)} \leq .... \leq U_{(n)}$ are the order statistics based on $U_1,...,U_n$. The DBEL approach is a technique to approximate values of $f_i$, $i=1,...,n$, via maximization of $DBEL_f$ given a constraint related to an empirical version of the density property $\int f(u)du = 1$. The DBEL testing approach revolves around exact test statistics which are independent of underlying data distributions under $H_0$. Recent developments of the DBEL techniques can be found in various statistical publications.[26-29]

Taking into account the setting of Section 2.1, in the case of completely specified forms of the density functions, we have that the likelihood ratio test statistic based on $Z_1,....,Z_n$ is

$$LR = \frac{\prod_{i=1}^{n} f_{H_1}(Z_i)}{\prod_{i=1}^{n} f_{H_0}(Z_i)} = \frac{\prod_{j=1}^{n} f_{H_1}(Z_{(j)})}{\prod_{j=1}^{n} f_{H_0}(Z_{(j)})} = \frac{\prod_{j=1}^{n} f_{H_1,j}}{\prod_{j=1}^{n} f_{H_0,j}},$$

where $f_{H_k}(u)$ denotes the density function of $Z_1$ under $H_k$ ($k=0,1$) and $Z_{(1)} \leq \cdots \leq Z_{(n)}$ are the order statistics based on observations $Z_1,..., Z_n$. Since in practice the data distributions are unknown, the DBEL approach focuses on approximating the values of $f_{H_k}(Z_{(j)})$ ($k=0,1$ and $j=1,..., n$) via maximizing the likelihood $\prod_{j=1}^{n} f_{H_k}(Z_{(j)})$ provided that $f_{H_k}(Z_{(j)}),...,f_{H_k}(Z_{(n)})$ satisfy an empirical constraint that corresponds to $\int f_{H_k}(u)du = 1$ under $H_k$.[25,28] To formalize this constraint, we specify, $Z_{(r)} = Z_{(1)}$ if $r \leq 1$, and $Z_{(r)} = Z_{(n)}$ if $r \geq n$, and employ the result

$$\frac{1}{2m} \sum_{j=1}^{n} \int_{Z_{(j-m)}}^{Z_{(j+m)}} f_{H_1}(u) du = \int_{Z_{(1)}}^{Z_{(n)}} f_{H_1}(u) du - \sum_{i=1}^{m-i} \frac{(m-i)}{2m} \left( \int_{Z_{(n-i)}}^{Z_{(n-i+1)}} f_{H_1}(u) du + \int_{Z_{(i)}}^{Z_{(i+1)}} f_{H_1}(u) du \right),$$

for all integer $m \leq \dfrac{n}{2}$. (2.3)

(See Proposition 2.1 of Vexler and Gurevich.[25]) Since $\int_{Z_{(1)}}^{Z_{(n)}} f_{H_1}(u)\,du \leq \int_{-\infty}^{+\infty} f_{H_1}(u)\,du = 1$, Equation (2.3) implies the inequality $(1/2m)\sum_{j=1}^{n}\int_{Z_{(j-m)}}^{Z_{(j+m)}} f_{H_1}(u)\,du \leq 1$. In this case, one can expect that $(1/2m)\sum_{j=1}^{n}\int_{Z_{(j-m)}}^{Z_{(j+m)}} f_{H_1}(u)\,du \approx 1$ when $m/n \to 0$, as $m,n \to \infty$.[30] Let $\hat{F}_{H_0}$ and $\hat{F}_{H_1}$ denote estimators of $F_{H_0}$ and $F_{H_1}$, respectively. Using the approximate analog to the mean-value integration theorem, one can derive the following empirical approximations

$$\sum_{j=1}^{n}\int_{Z_{(j-m)}}^{Z_{(j+m)}} f_{H_1}(u)\,du = \sum_{j=1}^{n}\int_{Z_{(j-m)}}^{Z_{(j+m)}} \dfrac{f_{H_1}(u)}{f_{H_0}(u)} f_{H_0}(u)\,du \approx \sum_{j=1}^{n}\dfrac{f_{H_1,j}}{f_{H_0,j}}\int_{Z_{(j-m)}}^{Z_{(j+m)}} f_{H_0}(u)\,du,$$

$$\approx \sum_{j=1}^{n}\dfrac{f_{H_1,j}}{f_{H_0,j}}\left(F_{H_0}(Z_{(j+m)}) - F_{H_0}(Z_{(j-m)})\right),$$

$\int_{Z_{(1)}}^{Z_{(n)}} f_{H_1}(u)\,du \approx \hat{F}_{H_1}(Z_{(n)}) - \hat{F}_{H_1}(Z_{(1)})$, and

$$\left(\int_{Z_{(n-i)}}^{Z_{(n-i+1)}} f_{H_1}(u)\,du + \int_{Z_{(i)}}^{Z_{(i+1)}} f_{H_1}(u)\,du\right) \approx \left(\hat{F}_{H_1}(Z_{(n-i+1)}) - \hat{F}_{H_1}(Z_{(n-i)})\right) + \left(\hat{F}_{H_1}(Z_{(i+1)}) - \hat{F}_{H_1}(Z_{(i)})\right).$$

Defining $\hat{F}_{H_1}$ as the empirical distribution function, then we obtain the empirical version of (2.3)

$$\dfrac{1}{2m}\sum_{j=1}^{n}\dfrac{f_{H_1,j}}{f_{H_0,j}}\left(\hat{F}_{H_0}(Z_{(j+m)}) - \hat{F}_{H_0}(Z_{(j-m)})\right) = \left(1 - \dfrac{1}{n}\right) - \dfrac{(m-1)}{2n}.$$

In order to define $\hat{F}_{H_0}$, we apply the distribution-free estimation for a symmetric distribution proposed by Schuster.[31] Thus, we obtain the empirical constraint

$$\dfrac{1}{2m}\sum_{j=1}^{n}\dfrac{f_{H_1,j}}{f_{H_0,j}}\Delta_{jm} = 1 - \dfrac{(m+1)}{2n}, \tag{2.4}$$

where $\Delta_{jm} := (2n)^{-1} \sum_{i=1}^{n} \{I(Z_i \leq Z_{(j+m)}) + I(-Z_i \leq Z_{(j+m)}) - I(Z_i \leq Z_{(j-m)}) - I(-Z_i \leq Z_{(j-m)})\}$. In order to find values of $f_{H_1,j}$ that maximize the log-likelihood $\sum_{j=1}^{n} \log(f_{H_1,j})$ subject to constraint (2.4), we derive $\partial/\partial f_{H_1,i}$, $i=1,\ldots, n$, from the Lagrange function

$$\Lambda_\lambda = \sum_{j=1}^{n} \log(f_{H_1,j}) + \lambda \left(1 - \frac{m+1}{2n} - \frac{1}{2m} \sum_{j=1}^{n} \frac{f_{H_1,j}}{f_{H_0,j}} \Delta_{jm}\right),$$

where $\lambda$ is the Lagrange multiplier. Then it is clear that the equation $\partial\Lambda_\lambda/\partial f_{H_1,j} = 0$ provides

$$f_{H_1,j} = f_{H_0,j} \frac{2m\left(1-(m+1)(2n)^{-1}\right)}{n\Delta_{jm}}, \quad j=1,\ldots,n.$$

This implies that the empirical maximum likelihood approximation to the likelihood $\prod_{j=1}^{n} f_{H_1,j}$, under alternative hypothesis $H_1$, can be presented as

$$\prod_{j=1}^{n} f_{H_0,j} \frac{2m\left(1-(m+1)(2n)^{-1}\right)}{n\Delta_{jm}}.$$

Therefore, the approximate *LR* has the empirical form

$$V_{nm} = \prod_{j=1}^{n} \frac{2m\left(1-(m+1)(2n)^{-1}\right)}{n\Delta_{jm}}. \tag{2.5}$$

Note that the test statistic (2.5) has a structure similar to those of statistics based on sample entropy that are known to have asymptotic optimal properties. [32,33]

The performance of the statistic $V_{nm}$ strongly depends on the unknown value of the integer parameter *m*. In order to eliminate the dependence on the parameter *m*, we use the maximum likelihood based methods shown in Vexler and Gurevich [25] and Vexler et al. [34] Then we employ the maximum likelihood principle to propose the test statistic

$$V_n = \min_{a(n) \le m \le b(n)} \prod_{j=1}^{n} 2m\left(1-(m+1)(2n)^{-1}\right) \Big/ (n\Delta_{jm}), \tag{2.6}$$

where $a(n) = n^{0.5+\delta}$, $b(n) = \min(n^{1-\delta}, n/2)$, and $\delta \in (0, 0.25)$. Following the DBEL literature, [30] we choose $\delta = 0.1$ in practice and define $\Delta_{jm} = 1/n$, if $\Delta_{jm} = 0$.

***The sequential DBEL approach:*** Finally, to sequentially test hypothesis (2.1), we define the following stopping rule

$$\tau_1 = \min\{n : n \ge 1, \ \log(V_n) \ge c_{\alpha,N}\}, \tag{2.7}$$

where $V_n$ is defined in (2.6), $\alpha$ is the significance level and $c_{\alpha,N}$ is the critical value associated with $N$ and $\alpha$ satisfying $\Pr_{H_0}(\tau_1 \le N) = \alpha$. Note that we set $\log(V_n) = 0$, for $n$=1, 2, 3, since $n \ge 4$ is required to compute the statistic $\log(V_n)$. The proposed decision making policy regarding stopping rule (2.7) consists of the following algorithm: 1) if, for the first time, for some $n(\le N)$, $\log(V_n)$ exceeds $c_{\alpha,N}$, we have $\tau_1 = n$ and the experiment is terminated at that stage along with the decision to reject $H_0$; 2) if, no such $n$ exists, the null hypothesis is not rejected along with the termination of the experiment at the target maximum sample size $N$. We consider derivation of $c_{\alpha,N}$ values in Section 2.4.

Since the stopping rule $\tau_1$ is based on the statistic $V_n$ which approximates the optimal parametric LR, the proposed test procedure can be anticipated to be very efficient. This is empirically confirmed in Sections 3 and 4.

The following proposition demonstrates the consistency of the proposed test.

***Proposition 1.*** Under $H_0$, we have

$$Pr_{H_0}\left\{\max_{1 \le n \le N} \log(V_n) > N^\gamma\right\} \to 0 \text{ as } N \to \infty,$$

whereas, under $H_1$, we have

$$Pr_{H_1}\left\{\max_{1\leq n\leq N}\log(V_n) > N^{\gamma}\right\} \to 1 \text{ as } N \to \infty,$$

where $V_n$ is defined by (2.6) and $\gamma \in (0.75, 1)$.

**Proof.** The proof of this proposition is outlined in Appendix A.

The Type I error probability related to the proposed test procedure (2.7) is $Pr_{H_0}\{\tau_1 \leq N\} = Pr_{H_0}\left\{\max_{1\leq n\leq N}\log(V_n) > c_{\alpha,N}\right\}$. Assume $c_{\alpha,N} = O(N^{\gamma})$ then Proposition 1 provides that $Pr_{H_0}\left\{\max_{1\leq n\leq N}\log(V_n) > c_{\alpha,N}\right\} \to 0$ as $N \to \infty$. Then it is clear that, in order to satisfy $Pr_{H_0}\{\tau_1 \leq N\} = \alpha$, for a fixed value $\alpha$ and large values of $N$, $c_{\alpha,N}$ should be in an order of $o(N^{\gamma})$. Thus, the proposed test procedure is consistent, since $Pr_{H_1}\left\{\max_{1\leq n\leq N}\log(V_n) > c_{\alpha,N}\right\} \to 1$ as $N \to \infty$, when $c_{\alpha,N}$ has an order smaller than that of $N^{\gamma}$. (In this case, $Pr_{H_0}\left\{\max_{1\leq n\leq N}\log(V_n) > N^{\gamma}\right\}$ is asymptotically a lower bound for $Pr_{H_0}\left\{\max_{1\leq n\leq N}\log(V_n) > c_{\alpha,N}\right\}$, when $c_{\alpha,N} < N^{\gamma}$ as $N \gg 1$.)

## 2.4 Null distribution of $\tau_1$

In this section, we show that the proposed test statistic is distribution-free under $H_0$. We then present the critical values for the new test procedure. The stopping rule $\tau_1$ contains the DBEL test statistics $V_1, V_2, ...$, which by definitions (2.4) and (2.6) depend only on certain indicator functions. It turns out that the null distribution of the stopping rule $\tau_1$ is independent of the distribution of the observations $Z_1, Z_2, \ldots$. In order to explain this claim, we note that under $H_0$

$$I(Z_i \leq Z_j) = I\left[\Phi^{-1}\{F_{H_0}(Z_i)\} \leq \Phi^{-1}\{F_{H_0}(Z_j)\}\right] \text{ and}$$

$$I(-Z_i \leq Z_j) = I\left[\Phi^{-1}\{F_{H_0}(-Z_i)\} \leq \Phi^{-1}\{F_{H_0}(Z_j)\}\right] = I\left[\Phi^{-1}\{1 - F_{H_0}(Z_i)\} \leq \Phi^{-1}\{F_{H_0}(Z_j)\}\right]$$

$$= I\left[-\Phi^{-1}\{F_{H_0}(Z_i)\} \leq \Phi^{-1}\{F_{H_0}(Z_j)\}\right], \text{ for } i \neq j \in [1, N],$$

where $\Phi^{-1}(x)$ denotes the inverse function of the standard normal cumulative distribution function $\Phi(x)$. This fact implies the Type I error rate is

$$\Pr_{H_0}\left\{\max_{1 \leq n \leq N} \log(V_n) > c_{\alpha, N}\right\} = \Pr_{Z_1, \ldots, Z_N \sim Normal(0,1)}\left\{\max_{1 \leq n \leq N} \log(V_n) > c_{\alpha, N}\right\}.$$

Then it is clear that the proposed procedure (2.7) is exact. Let $DBTS_N = \max_{1 \leq n \leq N}\{\log(V_n)\}$. Thus, by definition (2.6), $c_{\alpha, N}$ is the upper $\alpha$-percentile of the distribution of $DBTS_N$ satisfying $\Pr_{H_0}\{DBTS_N \geq c_{\alpha, N}\} = \alpha$. In a similar manner to the computing scheme shown in the SSRT procedure, we tabulate the critical values of $c_{\alpha, N}$ for various choices of $N$ and $\alpha$ using the MC approach based on 25,000 generations of $Z_1, \ldots, Z_N$. The results are shown in Table 1.

**Table 1.**

*Remark 1*. The MC method is a well-known approach for obtaining accurate approximations of the critical values for exact tests. [35] Vexler et al. [36] proposed an approach to compute critical values of exact test procedures using tabulated critical values and MC simulations in a Bayesian manner. In this framework, tabulated critical values are considered as prior information and simulated MC observations are used as data.

## 3. MONTE CARLO STUDY

To evaluate the performance of the new testing strategy as compared to the classical SSRT procedure, we carried out an extensive MC study. We examined the ASN and the corresponding statistical powers of the considered procedures. Critical values of the tests were set at the 5% level of significance and the MC experiments were repeated 10,000 times in each scenario based on $N$=25, 50 and 75, respectively. According to the statistical literature, [30,37] we used the alternatives in the MC study following the scenarios: (1) constant shifts in location, e.g., $X_i \sim N(0,1)$ and $Y_i \sim N(0.5,1)$, $i$=1,…,$N$; (2) constant and nonconstant shifts, e.g., $X_i \sim N(0,1)$ and $Y_i \sim N(0.5, 2^2)$, $i$=1,…,$N$; (3) skewed alternatives, e.g., $X_i \sim LogN(1,1)$ and $Y_i \sim LogN(1, 0.5^2)$, $i$=1,…,$N$; and (4) nonconstant shifts, e.g., $X_i \sim Beta(0.7,1)$ and $Y_i \sim Exp(2)$, $i$=1,…,$N$. Note that the statistical literature expects that the classical SSRT, which is based on the Wilcoxon signed-rank statistic, will be very efficient in scenario (1). Regarding Scenario (3), one can remark that that many measurements of markers related to health and social science have been shown to follow lognormal distributions. [38] The corresponding MC results are presented in Table 2. (In the Supplementary Material (SM), we provide Table S1 with additional outputs of the MC study.)

In Scenarios (1-2), the SSRT demonstrates a slightly higher power than the new test. (In the SM, we also show several cases, when the SSRT slightly outperforms the proposed procedure.) This result is consistent with the MC evaluations of the DBEL and the Wilcoxon signed rank tests in retrospective settings when observations are normally distributed. [30] The ASNs of the proposed test are comparable with those of the SSRT. In Scenario (3), when the pre- and/or post- measurements are lognormally distributed, the proposed testing procedure substantially outperforms the SSRT in terms of the statistical power and the ASN in the considered alternatives for $N=50, 75$. Consider the case when $Z_i = X_i - Y_i$, $i=1,\ldots,N$, $X_i \sim LogN(0,1)$ and $Y_i \sim U(1,2)$, $i=1,\ldots,N$, with $N=50, 75$. In this scenario, the SSRT shows the powers of 0.22 and 0.28 with the corresponding ASNs of 45 and 64, respectively, whereas the proposed test provides the powers of 0.70 and 0.98 with corresponding ASNs of 40 and 44, respectively. In Scenario (4), the proposed testing procedure has better performance in terms of the statistical power and the ASN than the SSRT in most of the considered cases.

Thus, compared to the SSRT, the sequential DBEL test procedure shows higher power and relatively smaller ASNs in many cases of the considered alternatives.

**Table 2.**

## 4. APPLICATION TO THE VENTILATOR-ASSOCIATED PNEUMONIA (VAP) STUDY

The VAP data were produced in the course of an institutional study at the State University of New York at Buffalo, in which oral treatments were compared to investigate their effects on infection of patients' respiratory system in an ICU. The pathogenesis of pneumonia, including VAP, involves aspiration of bacteria from the oropharynx into the lung, and subsequent failure of host defenses to clear the bacteria resulting in development of lung infection. The major potential respiratory bacterial pathogens (PRPs) include *Staphylococcus aureus*, *Pseudomonas aeruginosa*, *Acinetobacter sps*. and enteric species. Prior biomedical studies have found the association between strains from bronchocopic cultures isolated at the time pneumonia was suspected and dental plaque/mucosa that is often colonized by PRPs. [39,40] Thus, improving oral hygiene in MV-ICU patients and reducing dental plaque load on teeth has the potential to reduce the risk of VAP. The trial sequentially enrolled and examined 83 patients who were admitted to a trauma ICU of the Erie County Medical Center (ECMC). During this study, pre- and post-CHX treatment measurements of amount of aggregated bacteria (S. aureus, P. aeruginosa, Acinetobacter sps., and enteric organisms) were recorded from each patient. Results of quantitative cultures were expressed as colony forming units (cfu) per ml. Figure 1 shows the histogram of post- and pre-measurement differences of the 83 paired data points. (In the SM, Figure S1 presents the scatterplot based on pre- and post-CHX treatment measurements.) Based on the Shapiro-Wilk test of normality we reject the normality assumption for the observed paired data points (*p-value* = 0.03). Retrospectively, we apply the Wilcoxon signed-rank test based on the total 83 paired data points, obtaining the corresponding p-value of 0.04 ($<0.05$). Thus, the Wilcoxon singed-rank test rejects the hypothesis that there is no difference between pre- and post-measurements of aggregated bacteria. This conclusion is coherent to available clinical trial results that demonstrated an effect of CHX on the prevalence of oropharyngeal colonization of respiratory bacterial pathogens. [40] We examine that the proposed sequential DBEL testing scheme and the conventional SSRT procedure can detect the treatment effect in a more efficient manner, in a sense that the sequential methods will provide significant testing results based on less than the total sample size of 83.

In accordance with Section 2, we use the MC method to compute 95% critical values 2.676 and 5.166 for the SSRT and the proposed method, respectively, using *N*=83. The proposed sequential DBEL method rejected the null hypothesis that there is no difference between pre- and post-measurements based on 50 observations. However, the SSRT failed to reject the null hypothesis using the total 83 observations.

In order to evaluate the robustness of the proposed approach, we conducted the following bootstrap-type analyses. The strategy is that a sample with size *N* (<83) was randomly selected from the 83 paired data points to calculate the stopping numbers based on the sample using the proposed method and the SSRT. We repeated this strategy 5,000 times calculating the ASNs and the frequencies that the proposed method and the SSRT reject the null hypothesis. Table 3 presents these results for the different maximum sample sizes of *N*=15, 25, 35, 50, 65, and 75. In this Bootstrap-type study, we notice that 1) the proposed method has substantially higher rejections rates of a false null hypothesis than those of the SSRT; 2) the proposed method consistently generates smaller ASNs than those of the SSRT, resulting in significant savings in sample sizes.

**Figure 1.**

**Table 3.**

## 5. CONCLUSIONS

We have developed a new nonparametric sequential technique for detecting treatment effects based on paired data. The proposed method employs the density-based EL methodology that approximates the optimal likelihood ratio statistic in a distribution-free fashion. In the real world study, the new test clearly outperforms the conventional SSRT. To the best of our knowledge, perhaps, this article presents a research that belongs to a first cohort of studies related to applications of the EL techniques in sequential manners. The method developed in this article can be extended to construct various nonparametric group sequential approaches. For example, for a fixed integer $k$, the stopping rule (2.7) can be modified to be based on statistic $\log(V_{kn})$. Further empirical and theoretical studies are needed to evaluate the proposed approach in this framework.

It is clear that in the distribution-free setting considered in this article, there are no most powerful statistical procedures. Therefore it is very important to consider and evaluate reasonably developed decision making policies in the context of detecting treatment effects in practice.

## APPENDIX A. PROOF OF PROPOSITION 1

The proof of Proposition 1 is relatively complicated, since, in general, we need to evaluate properties of the statistic $max_{1 \leq n \leq N} \, log \left( min_{a(n) \leq m \leq b(n)} \prod_{j=1}^{n} 2m / (n\Delta_{jm}) \right)$, whereas the published asymptotic DBEL results consider $log \left( min_{a(n) \leq m \leq b(n)} \prod_{j=1}^{n} 2m / (n\Delta_{jm}) \right)$-type constructions as $n \to \infty$. The online supplementary material provides technical details of examining the property

$$Pr_{H_k} \left[ \bigcap_{n \leq N} \left\{ log \left( min_{a(n) \leq m \leq b(n)} \prod_{j=1}^{n} \frac{2m}{n\Delta_{jm}} \right) < N^{\gamma} \right\} \right] \to 1-k, \quad k = 0,1, \gamma \in (0.75,\ 1), \text{ as } N \to \infty$$

of the corresponding joint probabilities.

## SUPPLEMENTARY MATERIALS

Detailed proof of the proposition presented in Section 2, the R codes implementing the proposed method, the additional MC results are available in the supplementary materials and the scatterplot that illustrates the data considered in Section 4.

## ACKNOWLEDGEMENTS

This research was supported by the National Institutes of Health (NIH) grant 1G13LM012241-01. We are grateful to the Editor, Associate Editor and reviewers for their helpful comments that led to a substantial improvement in this paper.

**Table 1.** The critical values $c_{\alpha,N}$ of the new test procedure at the significance levels $\alpha$'s.

| N/$\alpha$ | 0.010 | 0.015 | 0.020 | 0.025 | 0.040 | 0.050 | 0.060 | 0.100 | 0.150 | 0.200 | 0.300 |
|---|---|---|---|---|---|---|---|---|---|---|---|
| 5  | 3.514 | 3.514 | 3.514 | 3.514 | 3.514 | 3.514 | 3.514 | 2.854 | 2.557 | 1.873 | 1.276 |
| 10 | 5.667 | 4.949 | 4.949 | 4.841 | 4.288 | 4.081 | 3.963 | 3.412 | 2.854 | 2.557 | 1.873 |
| 15 | 6.043 | 5.667 | 5.178 | 4.949 | 4.447 | 4.288 | 4.081 | 3.514 | 3.007 | 2.854 | 2.259 |
| 20 | 6.194 | 5.716 | 5.476 | 5.048 | 4.723 | 4.288 | 4.288 | 3.514 | 3.228 | 2.854 | 2.469 |
| 25 | 6.345 | 5.716 | 5.667 | 5.312 | 4.890 | 4.554 | 4.288 | 3.724 | 3.412 | 2.938 | 2.557 |
| 30 | 6.473 | 5.897 | 5.667 | 5.420 | 4.890 | 4.660 | 4.318 | 3.884 | 3.438 | 3.045 | 2.659 |
| 35 | 6.478 | 5.983 | 5.716 | 5.457 | 4.949 | 4.766 | 4.447 | 3.963 | 3.514 | 3.161 | 2.786 |

| | | | | | | | | | | |
|---|---|---|---|---|---|---|---|---|---|---|
| 40 | 6.478 | 6.026 | 5.716 | 5.525 | 4.949 | 4.830 | 4.534 | 4.030 | 3.514 | 3.253 | 2.854 |
| 45 | 6.478 | 6.123 | 5.716 | 5.638 | 4.949 | 4.841 | 4.589 | 4.081 | 3.514 | 3.338 | 2.854 |
| 50 | 6.478 | 6.168 | 5.751 | 5.667 | 5.027 | 4.890 | 4.723 | 4.121 | 3.614 | 3.412 | 2.882 |
| 55 | 6.575 | 6.194 | 5.783 | 5.667 | 5.061 | 4.926 | 4.735 | 4.211 | 3.685 | 3.441 | 2.940 |
| 60 | 6.724 | 6.345 | 5.959 | 5.716 | 5.176 | 4.949 | 4.841 | 4.288 | 3.768 | 3.514 | 3.027 |
| 65 | 6.594 | 6.278 | 5.906 | 5.716 | 5.158 | 4.949 | 4.843 | 4.288 | 3.839 | 3.514 | 3.091 |
| 70 | 6.739 | 6.345 | 6.046 | 5.716 | 5.280 | 4.980 | 4.890 | 4.288 | 3.930 | 3.537 | 3.161 |
| 75 | 6.734 | 6.345 | 6.048 | 5.716 | 5.321 | 5.017 | 4.929 | 4.309 | 3.976 | 3.606 | 3.231 |

**Table 2**. The MC powers and ASNs of the proposed sequential DBEL (Seq_DBEL) test and the SSRT at the significance level $\alpha = 0.05$.

| Scenario | $F_X$ | $F_Y$ | N | Seq_DBEL Power | Seq_DBEL ASN | SSRT Power | SSRT ASN |
|---|---|---|---|---|---|---|---|
| (S1) | N(0,1) | N(0.5,1) | 25 | 0.256 | 22 | 0.297 | 22 |
| | | | 50 | 0.446 | 40 | 0.534 | 39 |
| | | | 75 | 0.660 | 52 | 0.712 | 50 |
| (S2) | N(0,1) | N(0.5, $2^2$) | 25 | 0.124 | 24 | 0.144 | 24 |
| | | | 50 | 0.197 | 45 | 0.240 | 45 |
| | | | 75 | 0.298 | 65 | 0.334 | 64 |
| (S3) | LogN(1,1) | LogN(1, $0.5^2$) | 25 | 0.073 | 24 | 0.066 | 24 |
| | | | 50 | 0.167 | 47 | 0.088 | 48 |
| | | | 75 | 0.386 | 66 | 0.113 | 71 |
| | LogN(0,1) | U(1,2) | 25 | 0.168 | 23 | 0.162 | 23 |
| | | | 50 | 0.698 | 40 | 0.222 | 45 |
| | | | 75 | 0.981 | 44 | 0.286 | 64 |
| | LogN(0, $3^2$) | Chisq(6) | 25 | 0.314 | 23 | 0.186 | 23 |
| | | | 50 | 0.962 | 32 | 0.245 | 44 |
| | | | 75 | 1 | 33 | 0.315 | 62 |
| | LogN(1,1) | Gamma(5,1) | 25 | 0.233 | 22 | 0.283 | 22 |
| | | | 50 | 0.561 | 39 | 0.459 | 40 |
| | | | 75 | 0.867 | 48 | 0.607 | 53 |
| (S4) | Exp(1) | Gamma(2,3) | 25 | 0.14 | 24 | 0.138 | 24 |
| | | | 50 | 0.284 | 44 | 0.25 | 45 |
| | | | 75 | 0.482 | 61 | 0.365 | 63 |
| | Gamma(5,1) | Gamma(1,1/5) | 25 | 0.089 | 24 | 0.101 | 24 |

|  |  | 50 | 0.183 | 46 | 0.126 | 47 |
|  |  | 75 | 0.408 | 65 | 0.158 | 69 |
| Exp(1) | U(-1,2) | 25 | 0.249 | 23 | 0.268 | 23 |
|  |  | 50 | 0.565 | 39 | 0.497 | 40 |
|  |  | 75 | 0.853 | 47 | 0.688 | 51 |
| Gamma(4,2) | Gamma(5,2) | 25 | 0.235 | 22 | 0.283 | 22 |
|  |  | 50 | 0.412 | 41 | 0.501 | 39 |
|  |  | 75 | 0.621 | 54 | 0.680 | 51 |
| Chisq(1) | Beta(0.7,1) | 25 | 0.265 | 23 | 0.225 | 23 |
|  |  | 50 | 0.788 | 37 | 0.430 | 41 |
|  |  | 75 | 0.993 | 40 | 0.626 | 54 |
| Gamma(2,1) | U(1,2) | 25 | 0.178 | 23 | 0.161 | 24 |
|  |  | 50 | 0.511 | 42 | 0.305 | 44 |
|  |  | 75 | 0.890 | 50 | 0.443 | 61 |

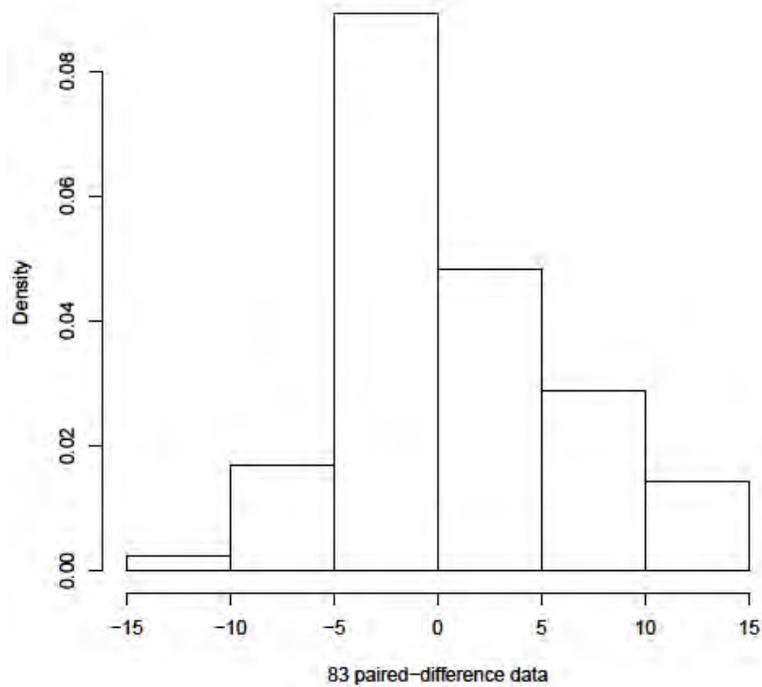

**Figure 1.** The histogram of the total 83 paired differences of pre- and post-CHX treatment measurements. The estimated mean, median and standard deviation of the 83 paired data points are 1.212, 0.002 and 5.036, respectively.

**Table 3.** The Bootstrap based rejection rates (RRs) and ASNs of the new test (Seq_DBEL) and the SSRT at the significance level of 0.05.

| $N$ | Seq_DBEL | | SSRT | |
|---|---|---|---|---|
| | RRs | ASNs | RRs | ASNs |
| 15 | 21.7% | 13 | 5.82% | 15 |
| 25 | 27.1% | 21 | 5.48% | 24 |
| 35 | 34.4% | 22 | 6.40% | 34 |
| 50 | 42.2% | 39 | 6.63% | 48 |
| 65 | 54.9% | 47 | 5.88% | 63 |
| 75 | 63.2% | 52 | 5.75% | 72 |

# A Sequential Density-Based Empirical Likelihood Ratio Test for Treatment Effects

Li Zou, Albert Vexler, Jihnhee Yu, and Hongzhi Wan

In this supplementary material, we provide the proof of Proposition 1, the R codes used in Sections 2-3 in this article, the additional MC results and the scatterplot that depicts the data considered in Section 4.

## 1. THE PROOF OF PROPOSITION 1.

***Proposition 1***. Under $H_0$, we have

$$\Pr\nolimits_{H_0}\left\{\max_{1\le n\le N}\log(V_n) > N^{\gamma}\right\} \to 0 \text{ as } N \to \infty,$$

whereas, under $H_1$, we have

$$\Pr\nolimits_{H_1}\left\{\max_{1\le n\le N}\log(V_n) > N^{\gamma}\right\} \to 1 \text{ as } N \to \infty,$$

where $V_n = \min_{a(n)\le m \le b(n)} \prod_{j=1}^{n} 2m\left(1-(m+1)(2n)^{-1}\right)\big/(n\Delta_{jm})$, $a(n) = n^{0.5+\delta}$, $b(n) = \min(n^{1-\delta}, 0.5n)$,

$$\Delta_{jm} = (2n)^{-1}\sum_{i=1}^{n}\left\{I\left(Z_i \le Z_{(j+m)}\right) + I\left(-Z_i \le Z_{(j+m)}\right) - I\left(Z_i \le Z_{(j-m)}\right) - I\left(-Z_i \le Z_{(j-m)}\right)\right\},$$

$\delta \in (0, 0.25)$, and $\gamma \in (0.75, 1)$.

*Proof.*

It is clear that the deterministic term $(m+1)/(2n)$ in the definition of $V_n$ quickly vanishes to zero. We simplify the explanation of the proof and represent $V_n$ in the form of $V_n = \min_{a(n)\le m \le b(n)} \prod_{j=1}^{n} 2m/(n\Delta_{jm})$. Toward this end, we will show that under $H_0$,

$$\Pr\nolimits_{H_0}\left\{\max_{1\leq n\leq N}\log\left(\min_{a(n)\leq m\leq b(n)}\prod_{j=1}^{n}\frac{2m}{n\Delta_{jm}}\right)>N^{\gamma}\right\}\to 0 \text{ as } N\to\infty,$$

and

$$\Pr\nolimits_{H_1}\left\{\max_{1\leq n\leq N}\log\left(\min_{a(n)\leq m\leq b(n)}\prod_{j=1}^{n}\frac{2m}{n\Delta_{jm}}\right)>N^{\gamma}\right\}\to 1 \text{ as } N\to\infty.$$

To prove that $\Pr\nolimits_{H_0}\left\{\max_{1\leq n\leq N}\log\left(\min_{a(n)\leq m\leq b(n)}\prod_{j=1}^{n}2m(n\Delta_{jm})^{-1}\right)>N^{\gamma}\right\}\to 0$ as $N\to\infty$, we use the trivial inequality $\min_{a(n)\leq m\leq b(n)}\prod_{j=1}^{n}2m(n\Delta_{jm})^{-1}\leq \prod_{j=1}^{n}2n^{1-\delta}(n\Delta_{jn^{1-\delta}})^{-1}$ for relatively large values of $n$ that implies

$$\Pr\nolimits_{H_0}\left\{\max_{1\leq n\leq N}\log\left(\min_{a(n)\leq m\leq b(n)}\prod_{j=1}^{n}\frac{2m}{n\Delta_{jm}}\right)>N^{\gamma}\right\}\leq \Pr\nolimits_{H_0}\left\{\max_{1\leq n\leq N}\log\left(\prod_{j=1}^{n}\frac{2n^{1-\delta}}{n\Delta_{jn^{1-\delta}}}\right)>N^{\gamma}\right\}. \quad (1)$$

Here we should note that the simple inequality

$$\max_{1\leq n\leq N}\log\left(\min_{a(n)\leq m\leq b(n)}\prod_{j=1}^{n}2m/(n\Delta_{jm})\right)\leq \max_{1\leq n\leq \exp(\log 2/\delta)}\log\left(\min_{a(n)\leq m\leq n/2}\prod_{j=1}^{n}2m/(n\Delta_{jm})\right)$$

$$+\max_{\exp(\log(2)/\delta)\leq n\leq N}\log\left(\min_{a(n)\leq m\leq n^{1-\delta}}\prod_{j=1}^{n}2m/(n\Delta_{jm})\right)$$

and the fact that $\Pr\left\{\max_{1\leq n\leq \exp(\log(2)/\delta)}\log\left(\min_{a(n)\leq m\leq n/2}\prod_{j=1}^{n}2m/(n\Delta_{jm})\right)>wN^{\gamma}\right\}\to 0$, for all fixed $0<w<1$ and $N\to\infty$, also justify the use of inequality (1).

In order to show that $\Pr_{H_0}\left\{\max_{1\leq n\leq N}\log\left(\prod_{j=1}^{n}2n^{1-\delta}\left(n\Delta_{jn^{1-\delta}}\right)^{-1}\right)>N^{\gamma}\right\}\to 0$, as $N\to\infty$, we use the following inequalities. Let $\gamma_1$ and $\gamma_2$ satisfy $0<\gamma_1<\gamma$ and $0<\gamma_2<\gamma$, we have

$$\Pr_{H_0}\left\{\max_{1\leq n\leq N}\log\left(\prod_{j=1}^{n}\frac{2n^{1-\delta}}{n\Delta_{jn^{1-\delta}}}\right)>N^{\gamma}\right\}=\Pr_{H_0}\left\{\max_{1\leq n\leq N}\sum_{j=1}^{n}\log\left(\frac{2n^{1-\delta}}{n\Delta_{jn^{1-\delta}}}\right)>N^{\gamma}\right\}$$

$$\leq \Pr_{H_0}\left\{\max_{1\leq n\leq N}\sum_{j=1}^{n^{1-\delta}}\log\left(\frac{2n^{1-\delta}}{n\Delta_{jn^{1-\delta}}}\right)+\max_{1\leq n\leq N}\sum_{j=n^{1-\delta}+1}^{n-n^{1-\delta}}\log\left(\frac{2n^{1-\delta}}{n\Delta_{jn^{1-\delta}}}\right)+\max_{1\leq n\leq N}\sum_{j=n-n^{1-\delta}+1}^{n}\log\left(\frac{2n^{1-\delta}}{n\Delta_{jn^{1-\delta}}}\right)>N^{\gamma}\right\}$$

$$=\Pr_{H_0}\left\{\max_{1\leq n\leq N}\sum_{j=1}^{n^{1-\delta}}\log\left(\frac{2n^{1-\delta}}{n\Delta_{jn^{1-\delta}}}\right)+\max_{1\leq n\leq N}\sum_{j=n^{1-\delta}+1}^{n-n^{1-\delta}}\log\left(\frac{2n^{1-\delta}}{n\Delta_{jn^{1-\delta}}}\right)+\max_{1\leq n\leq N}\sum_{j=n-n^{1-\delta}+1}^{n}\log\left(\frac{2n^{1-\delta}}{n\Delta_{jn^{1-\delta}}}\right)>N^{\gamma},\right.$$

$$\left.\max_{1\leq n\leq N}\sum_{j=1}^{n^{1-\delta}}\log\left(\frac{2n^{1-\delta}}{n\Delta_{jn^{1-\delta}}}\right)>N^{\gamma_1}\right\}$$

$$+\Pr_{H_0}\left\{\max_{1\leq n\leq N}\sum_{j=1}^{n^{1-\delta}}\log\left(\frac{2n^{1-\delta}}{n\Delta_{jn^{1-\delta}}}\right)+\max_{1\leq n\leq N}\sum_{j=n^{1-\delta}+1}^{n-n^{1-\delta}}\log\left(\frac{2n^{1-\delta}}{n\Delta_{jn^{1-\delta}}}\right)+\max_{1\leq n\leq N}\sum_{j=n-n^{1-\delta}+1}^{n}\log\left(\frac{2n^{1-\delta}}{n\Delta_{jn^{1-\delta}}}\right)>N^{\gamma},\right.$$

$$\left.\max_{1\leq n\leq N}\sum_{j=1}^{n^{1-\delta}}\log\left(\frac{2n^{1-\delta}}{n\Delta_{jn^{1-\delta}}}\right)\leq N^{\gamma_1}\right\}$$

$$\leq \Pr_{H_0}\left\{\max_{1\leq n\leq N}\sum_{j=1}^{n^{1-\delta}}\log\left(\frac{2n^{1-\delta}}{n\Delta_{jn^{1-\delta}}}\right)>N^{\gamma_1}\right\}+\Pr_{H_0}\left\{\max_{1\leq n\leq N}\sum_{j=n^{1-\delta}+1}^{n-n^{1-\delta}}\log\left(\frac{2n^{1-\delta}}{n\Delta_{jn^{1-\delta}}}\right)+\max_{1\leq n\leq N}\sum_{j=n-n^{1-\delta}+1}^{n}\log\left(\frac{2n^{1-\delta}}{n\Delta_{jn^{1-\delta}}}\right)>N^{\gamma}-N^{\gamma_1}\right\}$$

$$=\Pr_{H_0}\left\{\max_{1\leq n\leq N}\sum_{j=1}^{n^{1-\delta}}\log\left(\frac{2n^{1-\delta}}{n\Delta_{jn^{1-\delta}}}\right)>N^{\gamma_1}\right\}$$

$$+ \Pr_{H_0}\left\{\max_{1\le n\le N}\sum_{j=n^{1-\delta}+1}^{n-n^{1-\delta}}\log\left(\frac{2n^{1-\delta}}{n\Delta_{jn^{1-\delta}}}\right) + \max_{1\le n\le N}\sum_{j=n-n^{1-\delta}+1}^{n}\log\left(\frac{2n^{1-\delta}}{n\Delta_{jn^{1-\delta}}}\right) > N^\gamma - N^{\gamma_1}, \max_{1\le n\le N}\sum_{j=n^{1-\delta}+1}^{n-n^{1-\delta}}\log\left(\frac{2n^{1-\delta}}{n\Delta_{jn^{1-\delta}}}\right) > N^{\gamma_2}\right\}$$

$$+ \Pr_{H_0}\left\{\max_{1\le n\le N}\sum_{j=n^{1-\delta}+1}^{n-n^{1-\delta}}\log\left(\frac{2n^{1-\delta}}{n\Delta_{jn^{1-\delta}}}\right) + \max_{1\le n\le N}\sum_{j=n-n^{1-\delta}+1}^{n}\log\left(\frac{2n^{1-\delta}}{n\Delta_{jn^{1-\delta}}}\right) > N^\gamma - N^{\gamma_1}, \max_{1\le n\le N}\sum_{j=n^{1-\delta}+1}^{n-n^{1-\delta}}\log\left(\frac{2n^{1-\delta}}{n\Delta_{jn^{1-\delta}}}\right) \le N^{\gamma_2}\right\}$$

$$\le A + B + C, \tag{2}$$

where

$$A = \Pr_{H_0}\left\{\max_{1\le n\le N}\sum_{j=1}^{n^{1-\delta}}\log\left(\frac{2n^{1-\delta}}{n\Delta_{jn^{1-\delta}}}\right) > N^{\gamma_1}\right\},$$

$$B = \Pr_{H_0}\left\{\max_{1\le n\le N}\sum_{j=n^{1-\delta}+1}^{n-n^{1-\delta}}\log\left(\frac{2n^{1-\delta}}{n\Delta_{jn^{1-\delta}}}\right) > N^{\gamma_2}\right\},$$

$$C = \Pr_{H_0}\left\{\max_{1\le n\le N}\sum_{j=n-n^{1-\delta}+1}^{n}\log\left(\frac{2n^{1-\delta}}{n\Delta_{jn^{1-\delta}}}\right) > N^\gamma - N^{\gamma_1} - N^{\gamma_2}\right\}.$$

Consider item A. The statistic

$$\Delta_{jn^{1-\delta}} = \frac{1}{2n}\left(j + n^{1-\delta} - 1 + \sum_{i=1}^{n}I\left(Z_i \ge -Z_{(j+n^{1-\delta})}\right) - \sum_{i=1}^{n}I\left(Z_i \ge -Z_{(1)}\right)\right) \ge \frac{j + n^{1-\delta} - 1}{2n} \ge 0.5n^{-\delta},$$

for $j = 1,...,n^{1-\delta}$. Define $\beta_1$, satisfying $0 < \beta_1 < \gamma_1$, we obtain

$$A \le \Pr_{H_0}\left\{\max_{1\le n\le N^{\beta_1}}\sum_{j=1}^{n^{1-\delta}}\log\left(\frac{2n^{1-\delta}}{n\Delta_{jn^{1-\delta}}}\right) + \max_{N^{\beta_1}\le n\le N}\sum_{j=1}^{n^{1-\delta}}\log\left(\frac{2n^{1-\delta}}{n\Delta_{jn^{1-\delta}}}\right) > N^{\gamma_1}\right\}$$

$$\leq \Pr_{H_0}\left\{\max_{N^{\beta_1}\leq n\leq N}\sum_{j=1}^{n^{1-\delta}}\log\left(\frac{2n^{1-\delta}}{n\Delta_{jn^{1-\delta}}}\right) > N^{\gamma_1} - N^{\beta_1(1-\delta)}\log(4)\right\},$$

since $\Delta_{jn^{1-\delta}} \geq 0.5 n^{-\delta}$ is applied taking into account the term $\max_{1\leq n\leq N^{\beta_1}}\sum_{j=1}^{n^{1-\delta}}\log\left\{2n^{1-\delta}\left(n\Delta_{jn^{1-\delta}}\right)^{-1}\right\}$.

Define the function $D_n(x) = F_n(x) - F(x)$, where $F_n(x) = n^{-1}\sum_{i=1}^{n} I(Z_i \leq x)$ is the empirical distribution function. Then, for $1 \leq j \leq n^{1-\delta}$, we can rewrite $\Delta_{jn^{1-\delta}}$ in the form

$$\Delta_{jn^{1-\delta}} = \frac{j + n^{1-\delta} - 1}{2n} + \frac{1}{2}\left\{F_n\left(-Z_{(1)}\right) - F_n\left(-Z_{(j+n^{1-\delta})}\right)\right\},$$

where $F_n(-Z_{(1)}) = D_n(-Z_{(1)}) + F(-Z_{(1)})$ and $F_n\left(-Z_{(j+n^{1-\delta})}\right) = D_n\left(-Z_{(j+n^{1-\delta})}\right) + F\left(-Z_{(j+n^{1-\delta})}\right)$. The statement of $F(z) = 1 - F(-z)$ implies that, under $H_0$, we have

$$F_n(-Z_{(1)}) = D_n(-Z_{(1)}) + 1 - F(Z_{(1)}) = D_n(-Z_{(1)}) + 1 + F_n(Z_{(1)}) - F(Z_{(1)}) - F_n(Z_{(1)})$$

$$= D_n(-Z_{(1)}) + D_n(Z_{(1)}) + 1 - \frac{1}{n}.$$

In a similar manner to that shown above, we have

$$F_n\left(-Z_{(j+n^{1-\delta})}\right) = D_n\left(-Z_{(j+n^{1-\delta})}\right) + D_n\left(Z_{(j+n^{1-\delta})}\right) + 1 - n^{-1}\left(j + n^{1-\delta}\right).$$

Then

$$\Delta_{jn^{1-\delta}} = \frac{j + n^{1-\delta} - 1}{n} + \frac{1}{2}\left\{D_n(-Z_{(1)}) + D_n(Z_{(1)}) - D_n\left(-Z_{(j+n^{1-\delta})}\right) - D_n\left(Z_{(j+n^{1-\delta})}\right)\right\}.$$

Let $\beta_2$ satisfy $0 < 2\beta_2 < \beta_1$. We obtain

$$\text{Pr}_{H_0}\left\{\max_{N^{\beta_1}\leq n\leq N}\sum_{j=1}^{n^{1-\delta}}\log\left(\frac{2n^{1-\delta}}{n\Delta_{jn^{1-\delta}}}\right)>N^{\gamma_1}-N^{\beta_1(1-\delta)}\log(4)\right\}$$

$$=\text{Pr}_{H_0}\left\{\max_{N^{\beta_1}\leq n\leq N}\sum_{j=1}^{n^{1-\delta}}\log\left(\frac{2n^{1-\delta}}{n\Delta_{jn^{1-\delta}}}\right)>N^{\gamma_1}-N^{\beta_1(1-\delta)}\log(4),\max_{N^{\beta_1}\leq n\leq N}|D_n|>N^{-\beta_2}\right\}$$

$$+\text{Pr}_{H_0}\left\{\max_{N^{\beta_1}\leq n\leq N}\sum_{j=1}^{n^{1-\delta}}\log\left(\frac{2n^{1-\delta}}{n\Delta_{jn^{1-\delta}}}\right)>N^{\gamma_1}-N^{\beta_1(1-\delta)}\log(4),\max_{N^{\beta_1}\leq n\leq N}|D_n|\leq N^{-\beta_2}\right\}$$

$$\leq\text{Pr}_{H_0}\left(\max_{N^{\beta_1}\leq n\leq N}|D_n|>N^{-\beta_2}\right)+\text{Pr}_{H_0}\left\{\max_{N^{\beta_1}\leq n\leq N}\sum_{j=1}^{n^{1-\delta}}\log\left(\frac{2n^{1-\delta}}{n\Delta_{jn^{1-\delta}}}\right)>N^{\gamma_1}-N^{\beta_1(1-\delta)}\log(4),\max_{N^{\beta_1}\leq n\leq N}|D_n|\leq N^{-\beta_2}\right\}. \quad (3)$$

The proposition of Dvoretzky, Kiefer, and Wolfowitz [1, P.60] provides

$$\text{Pr}_{H_0}\left(\max_{N^{\beta_1}\leq n\leq N}|D_n|>N^{-\beta_2}\right)\leq\frac{C}{1-e^{-2N^{-2\beta_2}}}e^{-2N^{\beta_1-2\beta_2}},$$

where $C$ is a finite positive constant and $e^{-2N^{-2\beta_2}}\approx 1-2N^{-2\beta_2}$, as $N\to\infty$. These results yield the conclusion

$$\text{Pr}_{H_0}\left(\max_{N^{\beta_1}\leq n\leq N}|D_n|>N^{-\beta_2}\right)\to 0 \text{ as } N\to\infty, \quad (4)$$

where $0<2\beta_2<\beta_1$. Set up $\delta$ to satisfy $0<\delta<\beta_2$ and $1-\delta<\gamma_1$. In (3) under the event $\left\{\max_{N^{\beta_1}\leq n\leq N}|D_n|\leq N^{-\beta_2}\right\}$, we have

$$\max_{N^{\beta_1}\leq n\leq N}\sum_{j=1}^{n^{1-\delta}}\log\left(\frac{2n^{1-\delta}}{n\Delta_{jn^{1-\delta}}}\right)\leq -N^{1-\delta}\log\left(\frac{1}{2}-N^{\delta-\beta_2}\right)=O(N^{1-\delta}),$$

by the fact that $\Delta_{jn^{1-\delta}} \geq (j+n^{1-\delta}-1)n^{-1} - 2N^{-\beta_2} \geq n^{-\delta} - 2N^{-\beta_2}$. Considering (3) and $N^{\gamma_1} - N^{\beta_1(1-\delta)}\log(4) = O(N^{\gamma_1})$, we conclude that

$$\Pr_{H_0}\left\{\max_{N^{\beta_1} \leq n \leq N} \sum_{j=1}^{n^{1-\delta}} \log\left(\frac{2n^{1-\delta}}{n\Delta_{jn^{1-\delta}}}\right) > N^{\gamma_1} - N^{\beta_1(1-\delta)}\log(4), \max_{N^{\beta_1} \leq n \leq N}|D_n| \leq N^{-\beta_2}\right\} \to 0 \text{ as } N \to \infty. \quad (5)$$

By virtue of (4) and (5), we have

$$\Pr_{H_0}\left\{\max_{N^{\beta_1} \leq n \leq N} \sum_{j=1}^{n^{1-\delta}} \log\left(\frac{2n^{1-\delta}}{n\Delta_{jn^{1-\delta}}}\right) > N^{\gamma_1} - N^{\beta_1(1-\delta)}\log(4)\right\} \to 0 \text{ as } N \to \infty.$$

This leads to

$$A = \Pr_{H_0}\left\{\max_{1 \leq n \leq N} \sum_{j=1}^{n^{1-\delta}} \log\left(\frac{2n^{1-\delta}}{n\Delta_{jn^{1-\delta}}}\right) > N^{\gamma_1}\right\} \to 0 \text{ as } N \to \infty. \quad (6)$$

Now we consider the second item $B = \Pr_{H_0}\left\{\max_{1 \leq n \leq N} \sum_{j=n^{1-\delta}+1}^{n-n^{1-\delta}} \log\left(2n^{1-\delta}\left(n\Delta_{jn^{1-\delta}}\right)^{-1}\right) > N^{\gamma_2}\right\}$ at (2). The statistic

$$\Delta_{jn^{1-\delta}} = \frac{1}{2n}\left(2n^{1-\delta} + \sum_{i=1}^{n} I(Z_i \geq -Z_{(j+n^{1-\delta})}) - \sum_{i=1}^{n} I(Z_i \geq -Z_{(j-n^{1-\delta})})\right) \geq n^{-\delta},$$

for $j = n^{1-\delta}+1,\ldots,n-n^{1-\delta}$. Define $\beta_3$, satisfying $0 < \beta_3 < \gamma_2$, to obtain

$$B \leq \Pr_{H_0}\left\{\max_{1 \leq n \leq N^{\beta_3}} \sum_{j=n^{1-\delta}+1}^{n-n^{1-\delta}} \log\left(\frac{2n^{1-\delta}}{n\Delta_{jn^{1-\delta}}}\right) + \max_{N^{\beta_3} \leq n \leq N} \sum_{j=n^{1-\delta}+1}^{n-n^{1-\delta}} \log\left(\frac{2n^{1-\delta}}{n\Delta_{jn^{1-\delta}}}\right) > N^{\gamma_2}\right\}$$

$$\leq \Pr_{H_0}\left\{\max_{N^{\beta_3} \leq n \leq N} \sum_{j=n^{1-\delta}+1}^{n-n^{1-\delta}} \log\left(\frac{2n^{1-\delta}}{n\Delta_{jn^{1-\delta}}}\right) > N^{\gamma_2} - \left(N^{\beta_3} - 2N^{\beta_3(1-\delta)}\right)\log(2)\right\},$$

since $\Delta_{jn^{1-\delta}} \geq n^{-\delta}$ is applied taking into account the term $\max\limits_{1 \leq n \leq N^{\beta_3}} \sum\limits_{j=n^{1-\delta}+1}^{n-n^{1-\delta}} \log\left\{2n^{1-\delta}\left(n\Delta_{jn^{1-\delta}}\right)^{-1}\right\}$. In a similar manner to that we analyzed the item $\Delta_{jn^{1-\delta}}$, when $n^{1-\delta}+1 \leq j \leq n-n^{1-\delta}$, we obtain

$$\Delta_{jn^{1-\delta}} = 2n^{-\delta} + \frac{1}{2}\left\{D_n\left(-Z_{(j-n^{1-\delta})}\right) + D_n\left(Z_{(j-n^{1-\delta})}\right) - D_n\left(-Z_{(j+n^{1-\delta})}\right) - D_n\left(Z_{(j+n^{1-\delta})}\right)\right\}.$$

Define $\beta_4$, satisfying $0 < 2\beta_4 < \beta_3$, to obtain

$$\Pr\nolimits_{H_0}\left\{\max_{N^{\beta_3} \leq n \leq N} \sum_{j=n^{1-\delta}+1}^{n-n^{1-\delta}} \log\left(\frac{2n^{1-\delta}}{n\Delta_{jn^{1-\delta}}}\right) > N^{\gamma_2} - \left(N^{\beta_3} - 2N^{\beta_3(1-\delta)}\right)\log(2)\right\}$$

$$= \Pr\nolimits_{H_0}\left\{\max_{N^{\beta_3} \leq n \leq N} \sum_{j=n^{1-\delta}+1}^{n-n^{1-\delta}} \log\left(\frac{2n^{1-\delta}}{n\Delta_{jn^{1-\delta}}}\right) > N^{\gamma_2} - \left(N^{\beta_3} - 2N^{\beta_3(1-\delta)}\right)\log(2),\ \max_{N^{\beta_3} \leq n \leq N}|D_n| > N^{-\beta_4}\right\}$$

$$+ \Pr\nolimits_{H_0}\left\{\max_{N^{\beta_3} \leq n \leq N} \sum_{j=n^{1-\delta}+1}^{n-n^{1-\delta}} \log\left(\frac{2n^{1-\delta}}{n\Delta_{jn^{1-\delta}}}\right) > N^{\gamma_2} - \left(N^{\beta_3} - 2N^{\beta_3(1-\delta)}\right)\log(2),\ \max_{N^{\beta_3} \leq n \leq N}|D_n| \leq N^{-\beta_4}\right\}$$

$$\leq \Pr\nolimits_{H_0}\left(\max_{N^{\beta_3} \leq n \leq N}|D_n| > N^{-\beta_4}\right)$$

$$+ \Pr\nolimits_{H_0}\left\{\max_{N^{\beta_3} \leq n \leq N} \sum_{j=n^{1-\delta}+1}^{n-n^{1-\delta}} \log\left(\frac{2n^{1-\delta}}{n\Delta_{jn^{1-\delta}}}\right) > N^{\gamma_2} - \left(N^{\beta_3} - 2N^{\beta_3(1-\delta)}\right)\log(2),\ \max_{N^{\beta_3} \leq n \leq N}|D_n| \leq N^{-\beta_4}\right\}. \tag{7}$$

The result of Dvoretzky, Kiefer, and Wolfowitz [1, P.60] provides

$$\Pr\nolimits_{H_0}\left(\max_{N^{\beta_3} \leq n \leq N}|D_n| > N^{-\beta_4}\right) \to 0 \text{ as } N \to \infty, \tag{8}$$

where $0 < 2\beta_4 < \beta_3$. Set up $\delta$ to satisfy $0 < \delta < \beta_4$ and $1 + \delta - \beta_4 < \gamma_2$. In (7), under the event $\left\{\max\limits_{N^{\beta_3} \leq n \leq N}|D_n| \leq N^{-\beta_4}\right\}$, we have

$$\max_{N^{\beta_3} \leq n \leq N} \sum_{j=n^{1-\delta}+1}^{n-n^{1-\delta}} \log\left(\frac{2n^{1-\delta}}{n\Delta_{jn^{1-\delta}}}\right) \leq -\left(N - 2N^{1-\delta}\right)\log\left(1 - N^{\delta-\beta_4}\right) = O\left(N^{1+\delta-\beta_4}\right),$$

since $\Delta_{jn^{1-\delta}} \geq 2n^{-\delta} - 2N^{-\beta_4}$. Considering (7), since $N^{\gamma_2} - \left(N^{\beta_3} - 2N^{\beta_3(1-\delta)}\right)\log(2) = O\left(N^{\gamma_2}\right)$, we conclude that

$$\Pr_{H_0}\left\{\max_{N^{\beta_3} \leq n \leq N} \sum_{j=n^{1-\delta}+1}^{n-n^{1-\delta}} \log\left(\frac{2n^{1-\delta}}{n\Delta_{jn^{1-\delta}}}\right) > N^{\gamma_2} - \left(N^{\beta_3} - 2N^{\beta_3(1-\delta)}\right)\log(2), \max_{N^{\beta_3} \leq n \leq N}|D_n| \leq N^{-\beta_4}\right\} \to 0, \quad (9)$$

as $N \to \infty$. By virtue of (8) and (9), we have

$$\Pr_{H_0}\left\{\max_{N^{\beta_3} \leq n \leq N} \sum_{j=n^{1-\delta}+1}^{n-n^{1-\delta}} \log\left(\frac{2n^{1-\delta}}{n\Delta_{jn^{1-\delta}}}\right) > N^{\gamma_2} - \left(N^{\beta_3} - 2N^{\beta_3(1-\delta)}\right)\log(2)\right\} \to 0 \text{ as } N \to \infty.$$

This leads to

$$B = \Pr_{H_0}\left\{\max_{1 \leq n \leq N} \sum_{j=n^{1-\delta}+1}^{n-n^{1-\delta}} \log\left(\frac{2n^{1-\delta}}{n\Delta_{jn^{1-\delta}}}\right) > N^{\gamma_2}\right\} \to 0 \text{ as } N \to \infty. \quad (10)$$

Now we consider the third item $C = \Pr_{H_0}\left\{\max_{1 \leq n \leq N} \sum_{j=n-n^{1-\delta}+1}^{n} \log\left(2n^{1-\delta}\left(n\Delta_{jn^{1-\delta}}\right)^{-1}\right) > N^{\gamma} - N^{\gamma_1} - N^{\gamma_2}\right\}$ at (2).

The statistic

$$\Delta_{jn^{1-\delta}} = \frac{1}{2n}\left(n + n^{1-\delta} - j + \sum_{i=1}^{n} I\left(Z_i \geq -Z_{(n)}\right) - \sum_{i=1}^{n} I\left(Z_i \geq -Z_{(j-n^{1-\delta})}\right)\right) \geq \frac{n + n^{1-\delta} - j}{2n} \geq 0.5n^{-\delta},$$

for $j = n - n^{1-\delta} + 1, \ldots, n$. Define $\beta_5$, satisfying $0 < \beta_5 < \gamma$, to obtain

$$C \leq \Pr_{H_0}\left\{\max_{1 \leq n \leq N^{\beta_5}} \sum_{j=n-n^{1-\delta}+1}^{n} \log\left(\frac{2n^{1-\delta}}{n\Delta_{jn^{1-\delta}}}\right) + \max_{N^{\beta_5} \leq n \leq N} \sum_{j=n-n^{1-\delta}+1}^{n} \log\left(\frac{2n^{1-\delta}}{n\Delta_{jn^{1-\delta}}}\right) > N^{\gamma} - N^{\gamma_1} - N^{\gamma_2}\right\}$$

$$\leq \Pr_{H_0}\left\{\max_{N^{\beta_5}\leq n\leq N}\sum_{j=n-n^{1-\delta}+1}^{n}\log\left(\frac{2n^{1-\delta}}{n\Delta_{jn^{1-\delta}}}\right) > N^{\gamma} - N^{\gamma_1} - N^{\gamma_2} - N^{\beta_5(1-\delta)}\log(4)\right\},$$

since $\Delta_{jn^{1-\delta}} \geq 0.5n^{-\delta}$ is applied taking into account $\max_{1\leq n\leq N^{\beta_5}}\sum_{j=n-n^{1-\delta}+1}^{n}\log\left(2n^{1-\delta}/\left(n\Delta_{jn^{1-\delta}}\right)\right)$. In a similar manner to that we analyzed the item $\Delta_{jn^{1-\delta}}$, when $n - n^{1-\delta} + 1 \leq j \leq n$, we obtain

$$\Delta_{jn^{1-\delta}} = \frac{n + n^{1-\delta} - j}{n} + \frac{1}{2}\left\{D_n\left(-Z_{(j-n^{1-\delta})}\right) + D_n\left(Z_{(j-n^{1-\delta})}\right) - D_n\left(-Z_{(n)}\right) - D_n\left(Z_{(n)}\right)\right\}.$$

Define $\beta_6$, satisfying $0 < 2\beta_6 < \beta_5$, to obtain

$$\Pr_{H_0}\left\{\max_{N^{\beta_5}\leq n\leq N}\sum_{j=n-n^{1-\delta}+1}^{n}\log\left(\frac{2n^{1-\delta}}{n\Delta_{jn^{1-\delta}}}\right) > N^{\gamma} - N^{\gamma_1} - N^{\gamma_2} - N^{\beta_5(1-\delta)}\log(4)\right\}$$

$$= \Pr_{H_0}\left\{\max_{N^{\beta_5}\leq n\leq N}\sum_{j=n-n^{1-\delta}+1}^{n}\log\left(\frac{2n^{1-\delta}}{n\Delta_{jn^{1-\delta}}}\right) > N^{\gamma} - N^{\gamma_1} - N^{\gamma_2} - N^{\beta_5(1-\delta)}\log(4), \max_{N^{\beta_5}\leq n\leq N}|D_n| > N^{-\beta_6}\right\}$$

$$+ \Pr_{H_0}\left\{\max_{N^{\beta_5}\leq n\leq N}\sum_{j=n-n^{1-\delta}+1}^{n}\log\left(\frac{2n^{1-\delta}}{n\Delta_{jn^{1-\delta}}}\right) > N^{\gamma} - N^{\gamma_1} - N^{\gamma_2} - N^{\beta_5(1-\delta)}\log(4), \max_{N^{\beta_5}\leq n\leq N}|D_n| \leq N^{-\beta_6}\right\}$$

$$\leq \Pr_{H_0}\left(\max_{N^{\beta_5}\leq n\leq N}|D_n| > N^{-\beta_6}\right)$$

$$+ \Pr_{H_0}\left\{\max_{N^{\beta_5}\leq n\leq N}\sum_{j=n-n^{1-\delta}+1}^{n}\log\left(\frac{2n^{1-\delta}}{n\Delta_{jn^{1-\delta}}}\right) > N^{\gamma} - N^{\gamma_1} - N^{\gamma_2} - N^{\beta_5(1-\delta)}\log(4), \max_{N^{\beta_5}\leq n\leq N}|D_n| \leq N^{-\beta_6}\right\}. \quad (11)$$

The result of Dvoretzky, Kiefer, and Wolfowitz [1, P.60] provides

$$\Pr\nolimits_{H_0}\left(\max_{N^{\beta_5}\leq n\leq N}|D_n|>N^{-\beta_6}\right)\to 0 \text{ as } N\to\infty, \tag{12}$$

where $0<2\beta_6<\beta_5$. Set up $\delta$ to satisfy $0<\delta<\beta_6$ and $1-\delta<\gamma$. In (11), under the event $\left\{\max_{N^{\beta_5}\leq n\leq N}|D_n|\leq N^{-\beta_6}\right\}$, we have

$$\max_{N^{\beta_5}\leq n\leq N}\sum_{j=n-n^{1-\delta}+1}^{n}\log\left(\frac{2n^{1-\delta}}{n\Delta_{jn^{1-\delta}}}\right)\leq -N^{1-\delta}\log\left(\frac{1}{2}-N^{\delta-\beta_6}\right)=O(N^{1-\delta}),$$

since $\Delta_{jn^{1-\delta}}\geq (n+n^{1-\delta}-j)n^{-1}-2N^{-\beta_6}\geq n^{-\delta}-2N^{-\beta_6}$. Considering (11), since $N^{\gamma}-N^{\gamma_1}-N^{\gamma_2}-N^{\beta_5(1-\delta)}\log(4)=O(N^{\gamma})$, we conclude that

$$\Pr\nolimits_{H_0}\left\{\max_{N^{\beta_5}\leq n\leq N}\sum_{j=n-n^{1-\delta}+1}^{n}\log\left(\frac{2n^{1-\delta}}{n\Delta_{jn^{1-\delta}}}\right)>N^{\gamma}-N^{\gamma_1}-N^{\gamma_2}-N^{\beta_5(1-\delta)}\log(4),\right.$$

$$\left.\max_{N^{\beta_5}\leq n\leq N}|D_n|\leq N^{-\beta_6}\right\}\to 0 \text{ as } N\to\infty. \tag{13}$$

By virtue of (12) and (13), we have

$$\Pr\nolimits_{H_0}\left\{\max_{N^{\beta_5}\leq n\leq N}\sum_{j=n-n^{1-\delta}+1}^{n}\log\left(\frac{2n^{1-\delta}}{n\Delta_{jn^{1-\delta}}}\right)>N^{\gamma}-N^{\gamma_1}-N^{\gamma_2}-N^{\beta_5(1-\delta)}\log(4)\right\}\to 0 \text{ as } N\to\infty.$$

This leads to

$$C=\Pr\nolimits_{H_0}\left\{\max_{1\leq n\leq N}\sum_{j=n-n^{1-\delta}+1}^{n}\log\left(\frac{2n^{1-\delta}}{n\Delta_{jn^{1-\delta}}}\right)>N^{\gamma}-N^{\gamma_1}-N^{\gamma_2}\right\}\to 0 \text{ as } N\to\infty. \tag{14}$$

Applying the results (6), (10) and (14) to the Inequalities (1) and (2), we obtain

$$\Pr_{H_0}\left\{\max_{1\leq n\leq N}\log\left(\min_{a(n)\leq m\leq b(n)}\prod_{j=1}^{n}\frac{2m}{n\Delta_{jm}}\right)>N^{\gamma}\right\}\to 0 \text{ as } N\to\infty.$$

For example, one can set up $\delta = 0.15$, $\gamma = 0.95$, $\gamma_1 = \gamma_2 = 0.9$, $\beta_1 = 0.5$, $\beta_2 = 0.2$, $\beta_3 = 0.7$, $\beta_4 = 0.3$, $\beta_5 = 0.6$, $\beta_6 = 0.25$ to hold the requirements regarding the values of the parameters $\delta$, $\gamma$, $\gamma_1$, $\gamma_2$, $\beta_1$, $\beta_2$, $\beta_3$, $\beta_4$, $\beta_5$, and $\beta_6$ stated above. This completes the proposition 1's statement under $H_0$.

Consider $\Pr_{H_1}\left\{\max_{1\leq n\leq N}\log\left(\min_{a(n)\leq m\leq b(n)}\prod_{j=1}^{n}2m(n\Delta_{jm})^{-1}\right)>N^{\gamma}\right\}$ as $N\to\infty$. We have that

$$\max_{1\leq n\leq N}\log\left(\min_{a(n)\leq m\leq b(n)}\prod_{j=1}^{n}2m(n\Delta_{jm})^{-1}\right)\geq \log\left(\min_{a(N)\leq m\leq b(N)}\prod_{j=1}^{N}2m(N\Delta_{jm})^{-1}\right) \text{ implies}$$

$$\Pr_{H_1}\left\{\max_{1\leq n\leq N}\log\left(\min_{a(n)\leq m\leq b(n)}\prod_{j=1}^{n}\frac{2m}{n\Delta_{jm}}\right)>N^{\gamma}\right\}\geq \Pr_{H_1}\left\{\log\left(\min_{a(N)\leq m\leq b(N)}\prod_{j=1}^{N}\frac{2m}{N\Delta_{jm}}\right)>N^{\gamma}\right\}. \quad (15)$$

By virtue of Proposition 1 presented in Vexler et al.,[2] under $H_1$, we have

$$\frac{1}{N}\log\left(\min_{a(N)\leq m\leq b(N)}\prod_{j=1}^{N}\frac{2m}{N\Delta_{jm}}\right)\xrightarrow{p} -E\left\{\log\left(0.5+0.5\frac{f_{H_1}(-Z_1)}{f_{H_1}(Z_1)}\right)\right\} \text{ as } N\to\infty,$$

for all $0<\delta<1/4$, where $f_{H_1}(z)$ is the density function of $Z_1$ under $H_1$. Then

$$-E_{H_1}\left\{\log\left(0.5+0.5\frac{f_{H_1}(-Z_1)}{f_{H_1}(Z_1)}\right)\right\}>-\log\left\{0.5+0.5E_{H_1}\left(\frac{f_{H_1}(-Z_1)}{f_{H_1}(Z_1)}\right)\right\}\geq 0.$$

This leads to

$$\Pr_{H_1}\left\{\log\left(\min_{a(N)\leq m\leq b(N)}\prod_{j=1}^{N}\frac{2m}{N\Delta_{jm}}\right) > N^{\gamma}\right\} \to 1 \text{ as } N \to \infty. \tag{16}$$

Using the results of (15) and (16), we complete the proof.

## 2. R CODES USED IN SECTIONS 2-3

**2.1 R code to calculate critical values of the proposed test**

```
############################################################
######R code to calculate critical values of the proposed test#####
############################################################
# The number of the Monte Carlo simulations
MC<- 75000
# The fixed alpha value
alpha<- 0.05
# The maximum sample size N
N<- 15
# The value of parameter delta
delta<- 0.1
Vmc<- array()
for(mc in 1:MC){
x<- rnorm(N)
Vmc_n<-array()
# The loop is used to calculate values of the proposed test statistic for sample sizes n=4,…,N.
for(n in 4:N){
```

```r
Vnm<-array()

z<- x[1:n]

sz<- sort(z)

#Define values of m in (2.6)

m <- c(round(n^(delta+0.5))):min(c(round((n)^(1-delta)),round(n/2))))

delta_nm<-array()

#The function is used to calculate $\Delta_{jm}$ at (2.6)

    Vnm_function<- function(m){

    L<-c(1:n)- m

    LL<-replace(L, L <= 0, 1 )

    U<-c(1:n)+ m

    UU<-replace(U, U >= n, n)

    zL<-sz[LL]

    zU<-sz[UU]

    for (i in 1:n){

delta_nm[i]<- (sum(sz<=zU[i])+sum(-sz<=zU[i])-sum(sz<=zL[i])-sum(-sz<=zL[i]))/(2*n)}

delta_nm<- replace(delta_nm,delta_nm==0,1/n)

Vnm<- log(prod(m*(2*n-m-1)/(n^2*delta_nm)))

    return(Vnm)}

Vnm<- sapply(m,Vnm_function)

# Calculate the value of the test statistic $V_n$ in (2.6)

Vmc_n[n-3]<- min(Vnm)}

Vmc[mc]<- max(Vmc_n)}

# Obtain $c_{\alpha,N}$ of the proposed test

quantile(Vmc,1-alpha)
```

## 2.2 R code to calculate statistical powers and ASNs of the proposed test

```
#################X=U(1,2),Y=LogNormal(0,1)####################
# Define values of delta, N, number of the MC simulations, the critical value corresponding
#to alpha=0.05
delta<- 0.1
N<- 15
MC<- 100000
CV<- 4.288
N_stop<-array()
no_reject<-0
for(mc in 1:MC){
n=4
x=runif(N,min=1,max=2)-rlnorm(N,meanlog=0,sdlog=1)
while(n<=N){
Vnm<-array()
z<- x[1:n]
sz<- sort(z)
m <- c(round(n^(delta+0.5))):min(c(round((n)^(1-delta)),round(n/2))))
delta_nm<-array()
Vnm_function<- function(m){
      L<-c(1:n)- m
      LL<-replace(L, L <= 0, 1 )
      U<-c(1:n)+ m
      UU<-replace(U, U >= n, n)
      zL<-sz[LL]
```

```
        zU<-sz[UU]

        for (i in 1:n){

delta_nm[i]<- (sum(sz<=zU[i])+sum(-sz<=zU[i])-sum(sz<=zL[i])-sum(-sz<=zL[i]))/(2*n) }

delta_nm<- replace(delta_nm,delta_nm==0,1/n)

        Vnm<- log(prod(m*(2*n-m-1)/(n^2*delta_nm)))

return(Vnm)}

Vnm<- sapply(m,Vnm_function)

if(n<=N & min(Vnm)>=CV){

        N_stop[mc]<- n

        no_reject<- no_reject+1

        break}else if(n==N & min(Vnm)<CV){

N_stop[mc]=n}

n=n+1}}

#obtain the statistical power and ASN

print(c(no_reject/MC, mean(N_stop)))
```

## 3. THE ADDITIONAL MC RESULTS

**Table S1.** The MC powers and ASNs of the proposed sequential DBEL (Seq_DBEL) test and the SSRT at the significance level $\alpha = 0.05$.

|         |              |    | Seq_DBEL |     | SSRT  |     |
|---------|--------------|----|----------|-----|-------|-----|
| $F_X$   | $F_Y$        | N  | Power    | ASN | Power | ASN |
| N(0,1)  | Cauchy(1,1)  | 25 | 0.314    | 21  | 0.404 | 21  |
|         |              | 50 | 0.603    | 37  | 0.656 | 35  |
|         |              | 75 | 0.846    | 45  | 0.833 | 42  |
| Exp(1)  | LogN(0, $2^2$) | 25 | 0.287    | 23  | 0.269 | 23  |
|         |              | 50 | 0.704    | 37  | 0.502 | 40  |
|         |              | 75 | 0.951    | 42  | 0.701 | 51  |
| Beta(0.7,1) | Exp(2) | 25 | 0.059 | 24 | 0.054 | 25 |
| | | 50 | 0.087 | 48 | 0.062 | 49 |
| | | 75 | 0.215 | 70 | 0.073 | 72 |
| N(0,1) | U(-1,2) | 25 | 0.272 | 22 | 0.321 | 22 |
| | | 50 | 0.543 | 39 | 0.57 | 38 |
| | | 75 | 0.718 | 50 | 0.756 | 48 |
| U(1,2) | LogN(0,1) | 25 | 0.168 | 23 | 0.162 | 23 |
| | | 50 | 0.698 | 40 | 0.222 | 45 |
| | | 75 | 0.981 | 44 | 0.286 | 64 |
| Gamma(2,4) | Gamma(2,5) | 25 | 0.118 | 24 | 0.131 | 24 |
| | | 50 | 0.187 | 46 | 0.220 | 45 |
| | | 75 | 0.286 | 66 | 0.312 | 65 |
| Gamma(2,3) | Gamma(3,2) | 25 | 0.944 | 13 | 0.969 | 13 |
| | | 50 | 1 | 14 | 1 | 15 |
| | | 75 | 1 | 15 | 1 | 15 |
| Exp(1) | Beta(0.7,1) | 25 | 0.671 | 19 | 0.645 | 19 |
| | | 50 | 0.977 | 24 | 0.93 | 26 |
| | | 75 | 1 | 25 | 0.989 | 28 |
| Exp(1) | U(-1,2) | 25 | 0.249 | 23 | 0.268 | 23 |
| | | 50 | 0.565 | 39 | 0.497 | 40 |
| | | 75 | 0.853 | 47 | 0.688 | 51 |
| LogN(-0.5,1) | Beta(0.7,1) | 25 | 0.594 | 19 | 0.581 | 20 |
| | | 50 | 0.953 | 26 | 0.885 | 28 |
| | | 75 | 0.999 | 27 | 0.977 | 31 |
| Chisq(1) | U(1,2) | 25 | 0.605 | 19 | 0.617 | 18 |
| | | 50 | 0.986 | 24 | 0.877 | 26 |
| | | 75 | 1 | 25 | 0.968 | 29 |
| Gamma(0.9,1) | Beta(0.9,1) | 25 | 0.316 | 22 | 0.29 | 23 |
| | | 50 | 0.774 | 35 | 0.559 | 38 |
| | | 75 | 0.985 | 39 | 0.746 | 49 |
| Beta(0.7,1) | U(-1,1) | 25 | 0.698 | 18 | 0.691 | 18 |
| | | 50 | 0.975 | 23 | 0.952 | 24 |
| | | 75 | 0.999 | 24 | 0.994 | 26 |

**3.**

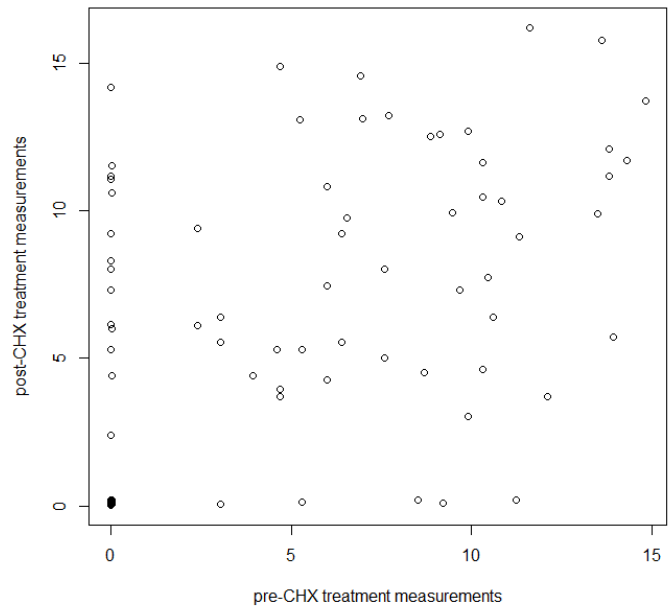

**Figure S1.** The scatterplot 'pre-CHX treatment measurements' vs. 'post-CHX treatment measurements'.